\documentclass[a4paper]{jpconf}
\usepackage{graphicx}
\usepackage{cite}
\usepackage{color}

\begin{document}

\title{High-Resolution Adiabatic Calorimetry of Supercooled Water}
\author{V.P.\,Voronov$^{1}$, V.E.\,Podnek$^{1, a}$ and M.A.\,Anisimov$^{2, b}$}
\address{$^{1}$Oil \& Gas Research Institute, Russian Academy of Sciences,
Moscow 119333, Russia}
\address{$^{2}$Department of Chemical and Biomolecular Engineering
and Institute for Physical Science and Technology, University of Maryland,
College Park, MD 20742,
U.S.A.}

\ead{podnek77@gmail.com$^a$, anisimov@umd.edu$^b$}

\begin{abstract}
Liquid water exhibits anomalous behavior in the supercooled region. A
 popular hypothesis to explain supercooled water's anomalies is the existence
 of a metastable liquid-liquid transition terminating at a critical point.
 The hypothesized phase transition is not directly accessible in a bulk
 experiment because it is expected to occur in ``no-man's" region below the kinetic stability
 limit of the liquid phase at about 235 K, the temperature of homogeneous
 ice formation. Therefore, verifications of this hypothesis are usually
 based on extrapolations from the experimentally accessible region.
 In this work, we present the results of high-resolution adiabatic
 calorimetry measurements of cold and supercooled liquid water in the range
 from 294 to 244 K, the lowest temperature of water's supercooling achieved
 so far in a bulk adiabatic-calorimetry experiment. The resolution of the
 measurements is also record-high, with the average statistical (random)
 error of about 0.1 \%. The data are consistent with adiabatic-calorimetry
 measurements of supercooled water earlier reported by Tombari {\it et
 al.} [Chem. Phys Lett. {\bf 300}, 749 (1999)] but significantly deviate from
 differential-scanning calorimetry measurements in emulsified water reported
 by Angell {\it et al.} [J. Phys. Chem. {\bf 86}, 998 (1982)] and by Archer and Carter
 [J. Phys. Chem. B {\bf 104}, 8563 (2000)]. Consequences of the new heat-capacity data
 in interpretation of the nature of water's anomalies are discussed.
\end{abstract}

\section{Introduction}

Liquid water exhibits anomalous behavior in the supercooled
region, as manifested by the growth of the isothermal
compressibility and isobaric heat capacity \cite{Speedy}. A
popular hypothesis to explain supercooled water's anomalies is the
existence of a metastable liquid-liquid transition terminating at
a critical point \cite{Pool, Mishima:1998, Debenedetti:1998,
Gallo}. The hypothesized phase transition is not directly
accessible in a bulk experiment because it is presumably located a
few degrees below the kinetic stability limit of the liquid phase
at about 235 K, the temperature of homogeneous ice formation
\cite{Debenedetti:1996, Caupin:2015}. Inevitably, verifications of
this hypothesis are usually based on extrapolations of the
anomalies observed in the experimentally accessible region
\cite{Fuentevilla, Holten:2012a, Holten:2012b,Holten:2014}.

The liquid-liquid coexistence, terminated at a critical point, has
been reported for some atomistic water models (see review
\cite{Gallo}), most notably in molecular simulations of the ST2
model of water \cite{Palmer:2014a,Palmer:2014b, Palmer:2018}, but
also in TIP4P/2005 model \cite{Yagasaki, Singh, Gonzalez}.
However, in simulation of some other models of water, such as mW
\cite{Holten:2013} no liquid-liquid transition was found. The
situation in real supercooled water is much less certain. While
the hypothesis of the metastable liquid-liquid separation has been
proved useful for building an equation of state and accurate
describing the thermodynamic anomalies in the experimentally
accessible region, these results cannot unambiguously confirm or
reject the underlying hypothesis. Thus, only further, more
comprehensive and accurate experiments on deeply supercooled water
could provide more robust and reliable information on the nature
of water's anomalies and the existence of a metastable critical
point.

One of the most informative thermodynamic property is the isobaric
heat capacity, $C_P$.  An anomaly of $C_P$ is an indication of
growing fluctuations of entropy, $\delta S$, since $\bigl<(\delta
S)^2\bigr> = k_B C_P$ (where $k_B$ is Boltzmann's constant). The
first adiabatic heat-capacity measurements in supercooled water
were reported by Anisimov {\it et al.} in 1972
\cite{Anisimov:1972}. The degree of supercooling was very modest,
$\sim$ 7 K, however, the data clearly demonstrated the growth of
the heat capacity in metastable liquid water upon supercooling. A
year later, in a breakthrough experiment, Angell {\it et al.}
\cite{Angell:1973} (further improved and summarized in ref.
\cite{Angell:1982}) were able to supercool water, emulsified in
heptane, down to 239 K. Using a differential scanning calorimeter
(DSC), they discovered a spectacular growth of the heat capacity
of metastable water upon supercooling. Some limited DSC data for
supercooled water were also reported by Bertolini {\it et al.} in
1985 \cite{Bertolini}. In 2000, Archer and Carter \cite{Archer},
also using a DSC and emulsified water, qualitatively confirmed the
results of Angell and co-workers, however, the discrepancy between
the two sets of data were not insignificant.

While a DSC method has certain advantages, being fast and
sensitive \cite{Wilhelm}, in investigations of phase transitions
and critical phenomena, unlike adiabatic calorimetry \cite{Thoen,
Anisimov, Anisimov:1991}, it may not be sufficiently accurate
because of long thermal equilibration. In 1999, Tombari {\it et
al.} \cite{Tombari} measured the isobaric heat capacity of bulk
supercooled water (down to 244.8 K) in 10 ml glass ampules with an
adiabatic calorimeter. The reported data significantly deviate
from those obtained by a DSC method.

In this work, we revisit this problem and present the results of
high-resolution adiabatic calorimetry measurements of cold and
supercooled liquid water in the range from 294 to 244 K, the
lowest temperature of water's supercooling achieved so far in a
bulk adiabatic-calorimetry experiment (0.8 K below the temperature
achieved by Tombari {\it et al.} \cite{Tombari}). While the
results are basically consistent with the data reported by Tombari
{\it et al.}, the resolution of our measurements is record-high,
with the statistical (random) error of about 0.1 \%. Remarkably,
the results of the both adiabatic calorimetry measurements
significantly deviate from differential-scanning calorimetry
measurements in water (emulsified in heptane) reported by Angell
{\it et al.} \cite{Angell:1982} and, even more significantly, from
the data reported by Archer and Carter \cite{Archer}. Consequences
of the new heat-capacity data in interpretation of the nature of
water's anomalies are discussed.

\section{EXPERIMENTAL TECHNIQUE AND MEASUREMENT PROCEDURE}

The isobaric heat-capacity measurements of cold and supercooled
water were performed with a high-resolution slow-scanning
adiabatic calorimeter.  This calorimeter was previously used for
studies of phase transitions and critical phenomena in liquid
crystals \cite{Voronov:1987}, polymers \cite{Voronov:1997}, fluids
and fluid mixtures \cite{Voronov:2003}, gas hydrates
\cite{Voronov:2007} and in porous media \cite{Entov,
Voronov:2016}. The adiabatic calorimeter enables one to measure
the heat capacity, heats of phase transitions, and estimate a
relaxation time of achieving thermal equilibrium. The measurements
can be carried out upon heating or cooling, in a temperature-step
regime or in a slow-scanning regime \cite{Thoen}. All the results
reported in this work were obtained in the scanning regime. The
chosen temperature rate was determined on required adiabatic
conditions.

The measurements were performed under saturated vapor pressure in
the range 0.05-0.7 kPa in sealed glass ampules containing 1 ml
dust-free, air-free water for medical injections (Microgen,
Irkutsk, Russia). In this range of saturated vapor pressure a
correction to the isobaric heat capacity of water is negligible
\cite{Anisimov:1991}. The samples were stored in a freezer for a
week. Only in a small fraction of the stored ampules (further used
for the measurements) water remained liquid at $\approx 245$ K
($\approx - 28$\,$^{\circ}$C).

\begin{figure}[h]
  \begin{center}
  \includegraphics[width=0.5\columnwidth]{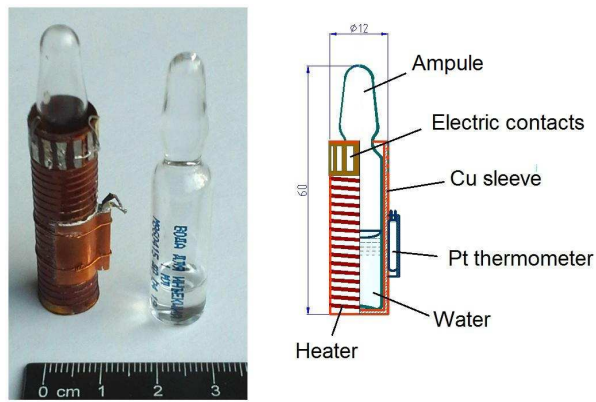}
  \caption{\label{fig1}Calorimetric cell with an ampule}
  \end{center}
\end{figure}

A custom-made calorimetric cell (a cylinder, 12 mm diameter) was
made of a 0.1 mm copper film (0.1 mm thickness). The ampules and
cell are shown in figure~\ref{fig1}. The temperature was measured
with a platinum-film (100 $\Omega$ resistance) thermometer
HEL-705-T-1-12, placed on the external surface of the cell. After
completing the measurements, used ampules were broken and dried,
then the mass and heat capacity of an empty ampule were measured
in order to obtain the mass and specific heat capacity of studied
water.

The heat-capacity measurements of supercooled water started to be
carried out in the regime of slow heating from the initial
temperature 243.25 K, at which the calorimeter was set for a few
hours to establish thermal equilibrium, while avoiding
crystallization. We note that the thermal equilibrium should not
be confused with global thermodynamic equilibrium because the
supercooled water is thermodynamically metastable.

\begin{figure}[h]
  \begin{center}
  \includegraphics[width=0.7\columnwidth]{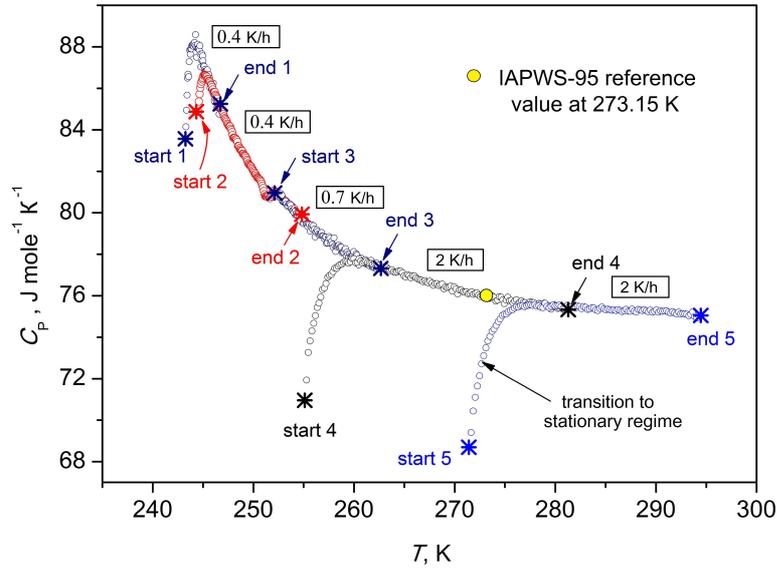}
  \caption{\label{fig2}Establishing stationary quasi-adiabatic
  regimes in slow-scanning adiabatic calorimetry measurements.
  The stationary adiabatic data obtained in different runs are overlapping.}
  \end{center}
\end{figure}

We divided the temperature range of the measurements in 5 narrow
intervals, as illustrated in figure~\ref{fig2}, for accurately
estimating the heat losses to be compensated to establish
adiabatic conditions. Within each interval the estimated heat loss
was linearized. The temperature scan for each interval started at
a ``star" point at a thermally nonequilibrium condition. The
equilibrium heat-capacity data were collected upon heating from a
low temperature after the system reached a stationary
quasi-adiabatic regime. If the stationary adiabatic runs obtained
in different intervals were overlapping, the data were considered
as reliable.

\section{RESULTS AND DISCUSSION
}

\begin{figure}[h]
  \begin{center}
  \includegraphics[width=0.7\columnwidth]{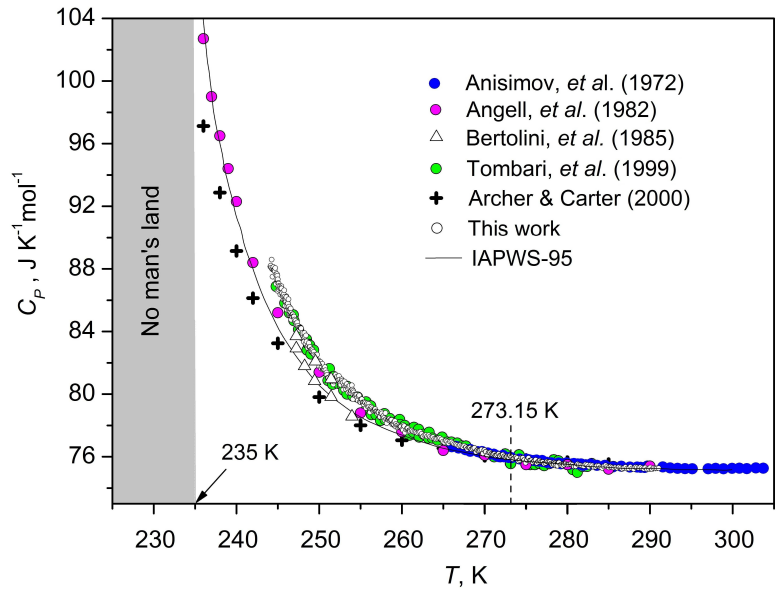}
  \caption{\label{fig3} Isobaric heat capacity (per mole)
of cold and supercooled water. Shown together with new data:
Anisimov {\it et al.} \cite{Anisimov:1972}, Angell {\it et al.}
\cite{Angell:1982}, Archer and Carter \cite{Archer}, Tombari {\it
et al.} \cite{Tombari} and Bertolini {\it et al.}
\cite{Bertolini}. Solid curve is IAPWS-95 (see ref.
\cite{Holten:2012a}).}
  \end{center}
\end{figure}

The results of our measurements, together with the data reported
earlier, are shown in figure~\ref{fig3}. All the data demonstrate
that the isobaric heat capacity of supercooled water exhibits a
striking anomaly with an apparent tendency to a ``divergence" or
to a sharp maximum at a temperature beyond the kinetic stability
limit of liquid water. However, a closer look reveals significant
deviations between the results obtained by DSC and
adiabatic-calorimeter measurements. The DSC data, especially those
obtained by Archer and Carter \cite{Archer}, are systematically
lower that the data obtained with slow-scanning adiabatic
calorimeters. Contrarily, the adiabatic measurements are mutually
consistent. However, the average statistical error (standard
deviation) in our measurements over the studied range of
temperatures is lower and the achieved supercooling is deeper. An
empirical approximation of the heat capacity anomaly, suggested by
the International Association for the Properties of Water and
Steam (IAPWS-95), is also shown in figure 4 (solid curve). This
approximation is, in fact, a fit to the data of Angell {\it et
al.} \cite{Angell:1982}. A comprehensive description of
supercooled water's anomalies, suggested by Holten {\it et al.}
\cite{Holten:2014} and in 2015 recommended by IAPWS as a guideline
for supercooled water (see ref. \cite{IAPWS}), predicts a
heat-capacity divergence at 228.2 K, while the $C_p$ curve in the
experimentally accessible region closely follows the shape of
IAPWS-95.

On the other hand, the results of our adiabatic measurements and
the data of Tombari et al. \cite{Tombari}; significantly deviate
from the DSC data and from approximations based on these data.

\begin{figure}[h]
  \begin{center}
  \includegraphics[width=0.7\columnwidth]{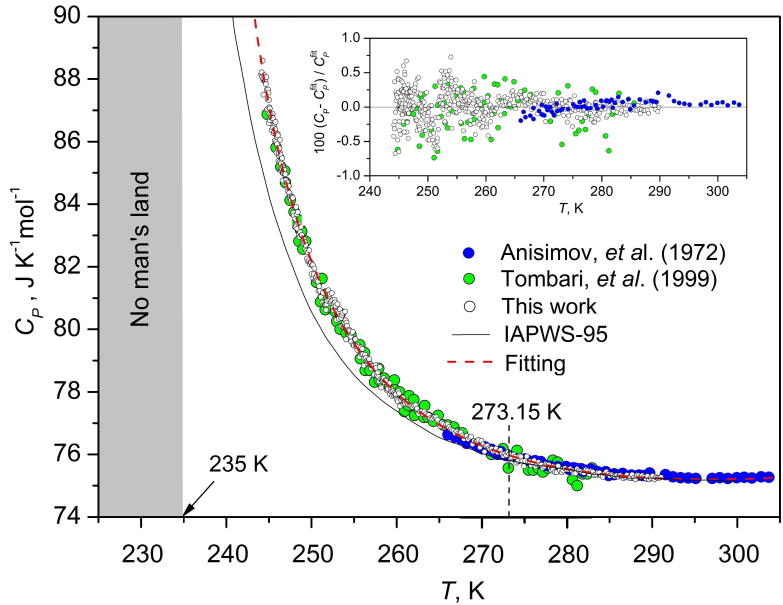}
\caption{\label{fig4} Approximation of the adiabatic heat-capacity
measurements in supercooled water by an empirical power law
(\ref{eq1}). Deviations of experimental data from the fit are
shown as the insert.}
  \end{center}
\end{figure}

To clarify the shape of the observed anomaly, we approximated all
available adiabatic calorimetry data obtained for cold and
supercooled water. The fitting results are demonstrated in
figure~\ref{fig4}. We tried an empirical power-law divergence at a
temperature $T^*$ (located below the kinetic stability limit of
liquid water):
\begin{eqnarray}
C_P/R =A\,\bigl((T-T^*)/ T^*\bigr)^{-n} + B\, (T/T^*) + C\,
,\label{eq1}
\end{eqnarray}
where $R$ is the universal gas constant. We obtain a surprisingly
good fit ($\chi^2/\text{DoF}=0.03$) with $A=0.16$, $B=1.68$,
$C=6.37$, and $n=0.89$. The extrapolated temperature of the
apparent divergence of the heat capacity, $T^*$=233.3 K appears to
be inside of ``no-man's" region, and higher than the temperature
of the heat-capacity maximum at about 228 K, predicted by a
two-structure equation of state, which estimates the hypothesized
liquid-liquid critical point at an elevated pressure
\cite{Holten:2012b} or a divergence at 228.2 K if the critical
point is at zero pressure \cite{Holten:2014}. However, this
apparent ``divergence" in no way can be regarded as an evidence of
the existence of a liquid-liquid transition and hypothesized
criticality in supercooled water because the extrapolation is made
from the region too far away from a possible heat-capacity
divergence or maximum. The analysis of supercooled water's
anomalies made with scaling \cite{Fuentevilla, Holten:2012a} and
two-structure models \cite{Holten:2012b,Holten:2014} show that an
effective power low as a reasonable approximation only in a narrow
range of temperatures. A more reliable temperature of the
heat-capacity divergence (if the measurements are assumed to be
performed along the critical isobar) or a temperature of the
heat-capacity maximum (if the isobar is assumed to be not
critical) could be obtained only from a comprehensive and
thermodynamically consistent analysis of all water's anomalies at
much deeper supercooling.

\section{CONCLUSION}

Adiabatic-calorimetry measurements under much deeper supercooling
are needed to unambiguously establish the shape of the
heat-capacity anomaly in supercooled water. To achieve this goal,
one may need to significantly reduce (down to milligrams) the
amount of waters samples. A collection of microcapillary samples
or long-lived emulsified water could be considered as promising
candidates. Upon further progress in calorimetric measurements,
the new data may require a reconsideration of the thermodynamic
description of supercooled waters anomalies, in particular, if the
concept of a hypothesized metastable liquid-liquid transition is
used for the construction of an equation of state.

\ack The research in Russia was carried out in the framework of
the Basic Research Program of the Russian Academy of Sciences. The
research at the University of Maryland, College Park was supported
by American Chemical Society Petroleum Research Fund (Grant No.
59434-ND6).

\section*{References}
\bibliographystyle{iopart-num}
\bibliography{Voronov2018}

\end{document}